\documentclass[pra, twocolumn, floatfix]{revtex4-2}
\usepackage{graphicx}
\usepackage{amsmath, amsfonts, amssymb, bm,color}

\def\ifa{Institute of Applied Physics, Moldova State University,
Academiei str. 5, MD-2028 Chi\c{s}in\u{a}u, Moldova}

\begin{document}
\title{Higher-order spatial photon interference versus dipole blockade effect}
\author{Arthur Rotari}
\affiliation{\ifa}
\author{Mihai A. Macovei}
\email{mihai.macovei@ifa.usm.md}
\affiliation{\ifa}
\date{\today}
\begin{abstract}
The steady-state quantum dynamics of three dipole-dipole coupled two-level emitters, fixed at the vertices of an equilateral 
triangle, and interacting via the environmental thermostat is investigated.  We have analytically obtained the populations 
of the involved three-atom cooperative states as well as of the second- and third-order spatial photon correlation functions 
of the light scattered by the few-qubit sample. As a consequence, we have demonstrated that this incoherently excited 
system spontaneously generates streams of single photons possessing sub-Poissonian photon statistics. In analogy to 
the dipole-dipole blockade, one may expect that at smaller inter particle distances, compared to the photon emission 
wavelength, the reported phenomenon has the same origin. However, we have shown that the quantum photon features 
are due to the interaction's nature of the few symmetrically arranged two-level emitters with the surrounding thermal 
reservoir. Respectively, at larger atomic intervals the effect occurs because of high-order spatial interference phenomena. 
Sub-wavelength interference fringes can be observed too, via measurements of spatial higher-order photon correlation 
functions.
\end{abstract}
\maketitle

\section{Introduction}
Generally, the light naturally emitted around us by various objects possesses classically incoherent properties.  However, 
the applications of modern technologies, including quantum ones, may require photon flows with quantum properties. 
Therefore, over the years, this topic has been widely investigated while the research leading to the generation of such 
photons is still being of great interest. Particularly, among those studies on quantum light generation is the work on 
photon anti-bunching in resonance fluorescence of sodium atoms continuously excited by a dye-laser beam \cite{ph_ab}. 
Squeezed light in single-atom resonance fluorescence occurs as well \cite{sq1}. Furthermore, the squeezing 
phenomenon can be even stronger in resonance fluorescence processes if $N$ atoms, distributed at regular positions, 
are involved \cite{sq2}. 

In order to generate single-, two- or multiple-photon quantum states, which are of practical interest too, the photon 
blockade effect may be involved. It is based on employing various Kerr-like non-linearities \cite{pbk0,pbk1}, strong 
qubit-cavity couplings \cite{pbk2,pbk3,pbk4} or coupled resonators \cite{pbk5,pbk6}, for instance. On the other side, 
when more than a single-qubit is considered, the strong dipole-dipole interactions among closely spaced few-level 
emitters cause shifting of the collective energy levels, resulting in off-resonant laser excitations. This leads to the so 
called dipole blockade phenomenon which can be used to inhibit excitation of higher collective states 
\cite{ddb1,ddb2,ddb3,ddb4}.

In this context, quantum features can be obtained even when the emitters interact with incoherent light or environmental 
thermal electromagnetic field reservoirs. For instance, entanglement between two arbitrary qubits can be generated if 
they interact with a common thermal bath \cite{enth1,enth2}, see also criteria for three-qubit entanglement \cite{tent}. 
A collection of strongly dipole-dipole interacting two-level emitters through the surrounding thermostat, in the Dicke 
limit, was shown to generate photons with sub-Poissonian statistics \cite{mma}. Remarkably, though counterintuitively, 
thermal light induces sub-wavelength interferences \cite{sqt1} which is a quantum effect as well, see also 
Refs.~\cite{boyd,shin}. Enhanced directional emission of spontaneous radiation can be produced also with statistically 
independent incoherent sources, via the measurement of higher-order correlation functions of the emitted radiation 
\cite{zanth}, see also \cite{tph1,tph2,zanth2}. It was shown there that in a linear chain of independent two-level emitters, 
the superradiance and the Hanbury Brown-Twiss effect basically complement each other. 

\begin{figure}[t]
\includegraphics[width =9cm]{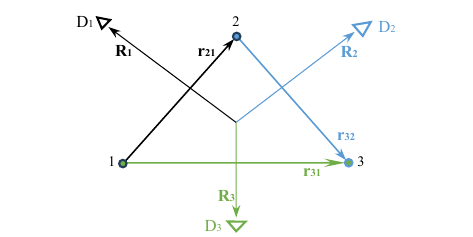}
\caption{\label{fig-1}
Schematic picture of the considered system. Three atoms, $j \in \{1,2,3\}$, are arranged to form an 
equilateral triangle. Detectors $D_{k}$, placed at ${\bf R_{k}}$, $k \in \{1,2,3\}$, detect the spontaneously 
scattered photons by the atomic sample. Respectively, ${\bf r_{jl}}$ are the inter atomic distance vectors 
and one assumes that $|{\bf R_{k}}| \gg |{\bf r_{jl}}|$.}
\end{figure}

Here, we theoretically investigate the interaction of a small system consisting of three quantum emitters (e.g. real 
or artificial atoms having two energy levels) which is mediated by the surrounding thermal electromagnetic field 
reservoir. The atomic subsystem is spatially arranged to form an equilateral triangle, see Figure~(\ref{fig-1}). 
Particularly, the focus is on the photon-scattering effects by the few-level quantum emitters which are excited 
incoherently by the thermostat, whereas the aim is being the quantum nature of these photons. In this context, 
we study the steady-state quantum dynamics of the considered sample and have obtained the atomic populations 
as well as the second- and the third-order spontaneously scattered photon correlation functions. The later are 
given as a function of detectors positions, respectively. We have found that for the employed geometrical arrangement 
of the atomic sample, the dipole-dipole interaction does not affect the steady-state properties of the photon 
scattering phenomenon. Hence, the sub-Poissonian photon statistics observed in the Dicke limit is rather due to 
the nature of the atoms-thermostat interaction and not because of the dipole-dipole blockade. Respectively, at 
larger inter atomic distances, the generated single photons stream possess quantum features too, however, in 
certain space directions. Furthermore, if two detectors are placed in symmetric positions, sub-wavelength 
interference fringes can be observed via measurements of the spatial second-order photon-photon correlation 
function. It happens because of the higher-order spatial quantum interference phenomena in thermal environments. 

Other relevant works have demonstrated that dipole-dipole interacting three-atom systems in external 
electromagnetic fields may break the dipole blockade \cite{berm} or can form three-body bound states 
\cite{mkf}, respectively. Three-qubit samples might be quite useful in context of quantum thermal transistors
\cite{qtht} and multifunctional quantum thermal devices \cite{qthm3} or in the interplay among the entanglement 
and non-periodic revival of spontaneous emission \cite{nrsp}. Three-body quantum antennas \cite{qant} or the 
triangular atomic mirrors \cite{trmir} are additional interesting issues of the few-qubit topic.

This paper is organized as follows. In Sec.~\ref{AT} we describe the analytical approach and the system of interest, 
while in Sec.~\ref{EqM} we discuss the steady-state population dynamics as well as the high-order photon correlation 
functions. Sec.~\ref{RD} presents and analyses the obtained results. The article concludes with a summary given in 
Sec.~\ref{sum}.

\section{The analytical treatment \label{AT}}
The Hamiltonian describing $N$ two-level emitters, all having identical transition frequencies $\omega_{0}$, and mutually 
interacting via their environmental electromagnetic thermal reservoir is given as follows \cite{gsag,wm,sczb,zficek,rewk}: 
$H=H_{0} + H_{i}$, where
\begin{eqnarray}
H_{0} &=& \sum_{k}\hbar \omega_{k}a^{\dagger}_{k}a_{k} + \sum^{N}_{j=1}\hbar\omega_{0}S_{zj}, \label{HH0}
\end{eqnarray}
and
\begin{eqnarray}
H_{i} =  i\sum_{k}\sum^{N}_{j=1}(\vec g_{k}\cdot \vec d_{j})
\bigl(a^{\dagger}_{k}S^{-}_{j}e^{-i\vec k\cdot \vec r_{j}}  - H.c. \bigr). \label{HHi}
\end{eqnarray}
Here, the first Hamiltonian, i.e. $H_{0}$, represents the free energy of the thermal reservoir as well as the corresponding 
free energy of the atomic subsystem, respectively. The Hamiltonian (\ref{HHi}) accounts for the interaction of the atomic 
subsystem with its surrounding thermal reservoir, where $\vec g_{k}$=$\sqrt{2\pi\hbar\omega_{k}/V}\vec e_{p}$ is the 
coupling strength among the few-level emitters and the thermal electromagnetic field modes, whereas $\vec e_{p}$ is 
the photon polarization vector with $p \in \{1,2\}$ and $V$ being the quantization volume. The atomic operators defined 
in the usual way, i.e., $S^{+}_{j} = |e\rangle_{j}{}_{j}\langle g|$ and $S^{-}_{j}=[S^{+}_{j}]^{\dagger}$, obey the 
usual commutation relations for su(2) algebra, namely, $[S^{+}_{j},S^{-}_{l}] =2S_{zj}\delta_{jl}$ and 
$[S_{zj},S^{\pm}_{l}]=\pm S^{\pm}_{j}\delta_{jl}$, where $S_{zj} = \bigl(|e\rangle_{j}{}_{j}\langle e|
-|g\rangle_{j}{}_{j}\langle g|\bigr)/2$ is the bare-state inversion operator of the $j$th qubit. Here, $|e\rangle_{j}$ and 
$|g\rangle_{j}$ are the excited and ground state of the emitter $j$, spatially localised at position $\vec r_{j}$, while 
$a^{\dagger}_{k}$ and $a_{k}$ are, respectively, the creation and the annihilation operators of the environmental 
electromagnetic thermal reservoir which satisfy the standard bosonic commutation relations, that is, 
$[a_{k}, a^{\dagger}_{k'}] = \delta_{kk'}$, and $[a_{k},a_{k'}]$ = $[a^{\dagger}_{k},a^{\dagger}_{k'}] = 0$ 
\cite{gsag,wm,sczb}.
\begin{figure}[t]
\includegraphics[width =10.5cm]{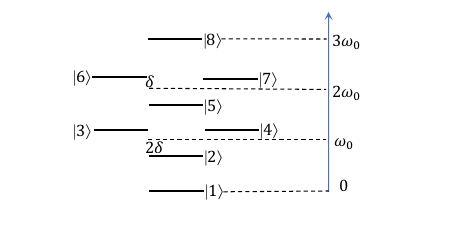}
\vspace{-1cm}
\caption{\label{fig-2}
The energy diagram of a spatially equidistant three-atom sample. The anti-symmetrical states $\{|6\rangle, |7\rangle \}$ 
and $\{|3\rangle, |4\rangle \}$ are shifted by the dipole-dipole coupling strength among the emitters, i.e. $+\delta$, 
whereas the symmetrical states $\{|5\rangle,|2\rangle\}$ are shifted by the doubled dipole-dipole coupling, $-2\delta$,
respectively.}
\end{figure}
 
Eliminating the thermal reservoir degrees of freedom, see e.g. \cite{sczb,zficek,rewk}, as usual in the Born-Markov 
approximations, one arrives at the following master equation describing the two-level ensemble in a thermal 
heat bath:
\begin{eqnarray}
\frac{d}{dt}\rho(t) &+& i\sum^{N}_{j\not=l}\delta_{ij}[S^{+}_{j}S^{-}_{l},\rho] = \nonumber \\
&-& \frac{\gamma}{2}\sum^{N}_{j,l=1}\chi_{jl}\bigl\{(1+\bar n)[S^{+}_{j},S^{-}_{l}\rho] 
\nonumber \\
&+& \bar n[S^{-}_{j},S^{+}_{l}\rho] \bigr\} + H.c.. \label{Mef}
\end{eqnarray}
Here,
\begin{eqnarray*}
\delta_{jl} &=& \frac{3\gamma}{4}\biggl\{\bigl[\cos^{2}\xi_{jl}-1\bigr]
\frac{\cos(\omega_{0}r_{jl}/c)}{(\omega_{0}r_{jl}/c)} \nonumber \\
&+& \bigl[1-3\cos^{2}\xi_{jl}\bigr]\biggl[\frac{\sin(\omega_{0}r_{jl}/c)}{(\omega_{0}r_{jl}/c)^{2}}
+ \frac{\cos(\omega_{0}r_{jl}/c)}{(\omega_{0}r_{jl}/c)^{3}}\biggr]\biggr\},
\end{eqnarray*}
are the vacuum-induced dipole-dipole interactions among any two atoms, separated by
$r_{jl}=|\vec r_{j}-\vec r_{l}|$, whereas
\begin{eqnarray*}
\chi_{jl} &=&\frac{3}{2}\biggl\{\bigl[1 - \cos^{2}\xi_{jl}\bigr]
\frac{\sin(\omega_{0}r_{jl}/c)}{(\omega_{0}r_{jl}/c)} \nonumber \\
&+& \bigl[1 - 3\cos^{2}\xi_{jl}\bigr]\biggl[\frac{\cos(\omega_{0}r_{jl}/c)}{(\omega_{0}r_{jl}/c)^{2}}
-\frac{\sin(\omega_{0}r_{jl}/c)}{(\omega_{0}r_{jl}/c)^{3}}\biggr]\biggr\},
\end{eqnarray*}
are the corresponding incoherent couplings contributing to the collective spontaneous decay \cite{zficek,rewk}, where $\xi_{jl}$ 
are the angles between the dipole moments $\vec d$, assumed identical for all atoms and parrallel, and $\vec r_{jl}$. The 
single emitter spontaneous decay rate is given by $\gamma=4d^{2}\omega^{3}_{0}/(3\hbar c^{3})$, whereas 
$\bar n=[\exp{(\hbar\omega_{0}/k_{B}T)}-1]^{-1}$ is the mean thermal photon number, at frequency 
$\omega_{0}$ and temperature $T$, with $k_{B}$ being the Bolzmann's constant, respectively. Actually, for small inter atomic
separations compared to the emission wavelength, i.e. $\omega_{0}r_{jl}/c\to 0$ in the Dicke limit, one has that $\chi_{jl} \to 1$, 
while $\delta_{jl}$ reduces to the static dipole-dipole interaction potential, i.e. $\delta_{jl} \to 3\gamma/[8(\omega_{0}r_{jl}/c)^{3}]$. 
Notice that $\chi_{jl}=\chi_{lj}$ and $\delta_{jl}=\delta_{lj}$, when $j\not=l$, and $\chi_{jj}=1$ while $\delta_{jj}$ reduces to 
a frequency shift, i.e. the Lamb shift, which is incorporated in the transition frequency $\omega_{0}$.

We next focus on a three-atom sample, where each emitter is being fixed at the vertices of an equilateral triangle. 
Therefore, the dipole-dipole coupling strengths are equal for any atomic pair, see Figure~(\ref{fig-1}). Hence, 
$\chi_{12}=\chi_{13}=\chi_{23}\equiv \chi$, while $\delta_{12} =\delta_{13}=\delta_{23}\equiv \delta$ and we assume 
that $\delta/\omega_{0} \to 0$. Although $\delta/\omega_{0} \to 0$, the dipole-dipole-coupling strength $\delta$ 
can be smaller and of the same order or larger than the single-emitter spontaneous decay rate $\gamma$, however, 
also $\gamma/\omega_{0} \to 0$. The three-atom collective Dicke states,  see e.g. \cite{thrF} and Figure~(\ref{fig-2}), 
are given then as follows: 
\begin{eqnarray}
|8\rangle &=&|e_{1}e_{2}e_{3}\rangle, \nonumber \\
|7\rangle&=&\bigl(|g_{1}e_{2}e_{3}\rangle - |e_{1}e_{2}g_{3}\rangle\bigr)/\sqrt{2}, 
\nonumber \\
|6\rangle &=&\bigl(-|g_{1}e_{2}e_{3}\rangle + 2|e_{1}g_{2}e_{3}\rangle - |e_{1}e_{2}g_{3}\rangle\bigr)/\sqrt{6},  \nonumber \\
|5\rangle &=&\bigl(|g_{1}e_{2}e_{3}\rangle + |e_{1}g_{2}e_{3}\rangle + |e_{1}e_{2}g_{3}\rangle\bigr)/\sqrt{3},  \nonumber \\
|4\rangle&=&\bigl(|e_{1}g_{2}g_{3}\rangle - |g_{1}g_{2}e_{3}\rangle\bigr)/\sqrt{2},  \nonumber \\
|3\rangle&=&\bigl(-|e_{1}g_{2}g_{3}\rangle + 2|g_{1}e_{2}g_{3}\rangle - |g_{1}g_{2}e_{3}\rangle\bigr)/\sqrt{6},  \nonumber \\
|2\rangle&=&\bigl(|e_{1}g_{2}g_{3}\rangle + |g_{1}e_{2}g_{3}\rangle + |g_{1}g_{2}e_{3}\rangle\bigr)/\sqrt{3},  \nonumber \\
|1\rangle &=& |g_{1}g_{2}g_{3}\rangle. 
\label{Dg3}
\end{eqnarray}
Taking into account that \cite{pbmm}: 
\begin{eqnarray}
S^{+}_{1} &=& (R_{41} + R_{87})/\sqrt{2} + (R_{21} - R_{64}- R_{73} \nonumber\\
&+& R_{85})/\sqrt{3} - ( R_{31} + R_{54} + R_{72} + R_{86})/\sqrt{6} \nonumber\\
&+& (R_{53} + R_{62})/\sqrt{18}+ 2(R_{52} - R_{63})/3, \nonumber  
\end{eqnarray}
\begin{eqnarray}
S^{+}_{2} &=& \sqrt{2/3}(R_{31} + R_{86}) + (R_{21} + R_{85})/\sqrt{3} \nonumber\\
&-& \sqrt{2}(R_{62}+ R_{53})/3 +  2(R_{63}/2 + R_{52})/3 \nonumber \\ 
&-& R_{74}, \nonumber \\ 
S^{+}_{3} &=& -( R_{41} + R_{87})/\sqrt{2} + (R_{21} + R_{64}+ R_{73} \nonumber \\ 
&+& R_{85})/\sqrt{3} + (R_{54} + R_{72} - R_{31} - R_{86})/\sqrt{6} \nonumber\\
&+& (R_{53} + R_{62})/\sqrt{18} + 2(R_{52} - R_{63})/3, \label{trSR}
\end{eqnarray}
with $S^{-}_{j}=[S^{+}_{j}]^{\dagger}$, $\{j \in 1,2,3\}$, the master equation (\ref{Mef}) takes the following form in the three-atom 
cooperative bases, namely,
\begin{eqnarray}
\frac{d}{dt}\rho(t)& + &\frac{i}{\hbar}[H,\rho]= -\frac{\gamma}{2}(1+\bar n)\bigl\{3(1+2\chi)[R^{+}_{\bar 1},R^{-}_{\bar 1}\rho] 
\nonumber \\
&+& 2(1-\chi)[R^{+}_{\bar 2},R^{-}_{\bar 2}\rho] + 6(1-\chi)[R^{+}_{\bar 3},R^{-}_{\bar 3}\rho] \bigr \} \nonumber \\
&-&\frac{\gamma}{2}\bar n\bigl\{3(1+2\chi)[R^{-}_{\bar 1},R^{+}_{\bar 1}\rho] - 2(1-\chi) \nonumber \\
&\times&[R^{-}_{\bar 2},R^{+}_{\bar 2}\rho] + 6(1-\chi)[R^{-}_{\bar 3},R^{+}_{\bar 3}\rho] \bigr \} + H.c., \nonumber \\
\label{MeqT}
\end{eqnarray}
where the three-atom collective operators are defined as $R_{\alpha\beta}=|\alpha\rangle\langle \beta|$, 
$\{\alpha, \beta \in 1, \cdots, 8\}$, satisfying the following commutation relations: 
$[R_{\alpha\beta},R_{\beta'\alpha'}]=R_{\alpha\alpha'}\delta_{\beta\beta'} - R_{\beta'\beta}\delta_{\alpha'\alpha}$. 
In Eq.~(\ref{MeqT}), $R^{+}_{\bar 1} = (R_{21} + R_{85})/\sqrt{3} + 2R_{52}/3-(R_{74} + R_{63})/3$, $R^{+}_{\bar 2} = (R_{41} 
+ R_{87})/\sqrt{2} - (R_{73} + R_{64})/\sqrt{3} - (R_{72} + R_{54})/\sqrt{6}$, and $R^{+}_{\bar 3} = (R_{62} + R_{53})/(3\sqrt{2}) 
- (R_{31} + R_{86})/\sqrt{6} + (R_{74} - R_{63})/3$, with $R^{-}_{\bar j}=[R^{+}_{\bar j}]^{\dagger}$, $\{\bar j \in \bar 1,\bar 2,\bar 3\}$, 
respectively. The Hamiltonian $H$, entering in Eq.~(\ref{MeqT}), is given by $H=H_{q} + H_{dd}$, where
\begin{eqnarray}
H_{q} &=& \hbar\omega_{0}\bigl(R_{22} + R_{33} + R_{44}\bigr) \nonumber \\
&+& 2\hbar\omega_{0}\bigl(R_{55} + R_{66} + R_{77}\bigr) + 3\hbar\omega_{0}R_{88}, 
\label{Hq}
\end{eqnarray}
while
\begin{eqnarray}
H_{dd}&=& -2\hbar\delta\bigl(R_{22} + R_{55}\bigr) \nonumber \\
&+& \hbar\delta\bigl(R_{33} + R_{44} + R_{66} + R_{77}\bigr). \label{Hdd}
\end{eqnarray}
One can observe that in the Dicke limit, when $\chi \to 1$, the master equation (\ref{MeqT}) reduces to
\begin{eqnarray}
\frac{d}{dt}\rho(t)& + &\frac{i}{\hbar}[H,\rho]= -\frac{9\gamma}{2}\bigl(1 + \bar n\bigr)\bigl[R^{+}_{\bar 1},R^{-}_{\bar 1}\rho\bigr] 
\nonumber \\
&-&\frac{9\gamma}{2}\bar n\bigl[R^{-}_{\bar 1},R^{+}_{\bar 1}\rho\bigr] + H.c., \label{MeqD}
\end{eqnarray}
meaning that only the symmetrical three-atom collective states are involved, namely, $|8\rangle$, $|5\rangle$, $|2\rangle$ and 
$|1\rangle$, see Figure~(\ref{fig-2}) and Exps.~(\ref{Dg3}). Indeed, there are totally $N+1 \equiv 3+1=4$ cooperative Dicke states
 in this particular case. The remaining anti-symmetrical three-atom collective states, i.e., $|3\rangle$, $|4\rangle$, $|6\rangle$ 
and $|7\rangle$ decouple from the interaction with the environmental thermal electromagnetic field reservoir and can be then 
omitted, in the Dicke limit. Thus, generally, the master equation (\ref{MeqT}) involves symmetrical as well as anti-symmetrical 
three-atom collective states, which are shifted proportional to the dipole-dipole coupling strength $\delta$, respectively, see 
Figure~(\ref{fig-2}).

In the following Section, we shall focus on the steady-state collective quantum dynamics of the three-qubit sample, arranged 
in an equilateral triangle configuration, where both the symmetrical as well as the anti-symmetrical cooperative states are 
being involved.
\section{Equations of motion and photon correlation functions \label{EqM}}
Using the master equation (\ref{MeqT}), one obtains the exact system of equations of motion for the mean values of the populations
in the symmetrical as well as anti-symmetrical three-atom collective states:
\begin{eqnarray}
\langle \dot R_{11}\rangle &=& -3\gamma\bar n\langle R_{11}\rangle + \gamma(1+2\chi)(1+\bar n)\langle R_{22}\rangle 
\nonumber \\
&+& \gamma(1-\chi)(1+\bar n)\langle R_{x}\rangle, \nonumber \\
\langle \dot R_{22}\rangle &=& \gamma\bar n(1+2\chi)\langle R_{11}\rangle - \gamma\bigl(1+2\chi 
\nonumber \\
&+&\bar n(3+4\chi)\bigr)\langle R_{22}\rangle +\frac{4\gamma}{3}(1+\bar n)(1+2\chi)\langle R_{55}\rangle 
\nonumber \\
&+& \frac{\gamma}{3}(1+\bar n)(1-\chi)\langle R_{y}\rangle, \nonumber \\
\langle \dot R_{55}\rangle &=& \frac{4\gamma}{3}\bar n(1+2\chi)\langle R_{22}\rangle  + \gamma(1+\bar n)(1+2\chi)\langle R_{88}\rangle
\nonumber \\
&-&\gamma\bigl(2(1+\chi) +\bar n(3+4\chi)\bigr)\langle R_{55}\rangle \nonumber \\
&+& \frac{\gamma}{3}\bar n(1-\chi)\langle R_{x}\rangle, \nonumber \\
\langle \dot R_{88}\rangle &=& \gamma\bar n(1+2\chi)\langle R_{55}\rangle - 3\gamma(1+\bar n)\langle R_{88}\rangle \nonumber \\
&+& \gamma\bar n(1-\chi)\langle R_{y}\rangle, \nonumber \\
\langle \dot R_{x}\rangle &=&2\gamma\bar n(1-\chi)\langle R_{11}\rangle  + \frac{2\gamma}{3}(1+\bar n)(1-\chi)\langle R_{55}\rangle
\nonumber \\
&-&\gamma\bigl(1 - \chi  + \bar n(3 - 2\chi)\bigr)\langle R_{x}\rangle + \frac{\gamma}{3}(1+\bar n) \nonumber \\
&\times&(5 - 2\chi)\langle R_{y}\rangle, \nonumber \\
\langle \dot R_{y}\rangle &=& \frac{2\gamma}{3}\bar n(1-\chi)\langle R_{22}\rangle + 2\gamma(1+\bar n)(1-\chi)\langle R_{88}\rangle 
\nonumber \\
&+& \frac{\gamma}{3}\bar n(5-2\chi)\langle R_{x}\rangle - \gamma\bigl(2-\chi  \nonumber \\
&+& \bar n(3-2\chi)\bigr)\langle R_{y}\rangle, \label{EqMt}
\end{eqnarray}
where the overdot means differentiation with respect to time, while 
\begin{eqnarray*}
\langle R_{x}\rangle = \langle R_{33}\rangle + \langle R_{44}\rangle, ~~{\rm and}~~
\langle R_{y}\rangle = \langle R_{66}\rangle + \langle R_{77}\rangle, 
\end{eqnarray*}
so that
\begin{eqnarray*}
\sum^{8}_{i=1}\langle R_{ii}\rangle=1.
\end{eqnarray*}
One can observe that the populations of the symmetrical and anti-symmetrical collective states will decouple in the Dicke limit, 
i.e. when $\chi = 1$. Additionally, for a symmetrical geometrical arrangement of three emitters, as it is the case here, the 
populations of the symmetrical states will not depend on the dipole-dipole coupling strength $\delta$, see Eqs.~(\ref{EqMt}). 

For later purposes, we write down the missing equations of motion connecting, respectively, the symmetrical and anti-symmetrical 
three-qubit decay channels, that is,
\begin{eqnarray}
\langle \dot R_{33}\rangle &=& \gamma\bar n(1-\chi)\langle R_{11}\rangle - \gamma\bigl(1-\chi +\bar n(3-2\chi)\bigr)\nonumber \\
&\times&\langle R_{33}\rangle +\frac{\gamma}{3}(1+\bar n)(1-\chi)\bigl(\langle R_{55}\rangle + 2\langle R_{77}\rangle \bigr) 
\nonumber \\
&+& \gamma(1+\bar n)\langle R_{66}\rangle - \frac{2\gamma}{3\sqrt{2}}(1+\bar n)(1-\chi)(\langle R_{65}\rangle 
\nonumber \\
&+& \langle R_{56}\rangle), \nonumber \\
\langle \dot R_{44}\rangle &=& \gamma\bar n(1-\chi)\langle R_{11}\rangle - \gamma\bigl(1-\chi +\bar n(3-2\chi)\bigr)
\nonumber \\
&\times&\langle R_{44}\rangle +\frac{\gamma}{3}(1+\bar n)(1-\chi)\bigl(\langle R_{55}\rangle + 2\langle R_{66}\rangle \bigr) 
\nonumber \\
&+& \gamma(1+\bar n)\langle R_{77}\rangle +\frac{2\gamma}{3\sqrt{2}}(1+\bar n)(1-\chi)(\langle R_{65}\rangle  \nonumber \\
&+& \langle R_{56}\rangle), 
\nonumber \\
\langle \dot R_{66}\rangle &=&\frac{\gamma\bar n}{3}(1-\chi)(\langle R_{22}\rangle + 2\langle R_{44}\rangle) - 
\gamma\bigl(2-\chi  \nonumber \\
&+&\bar n(3-2\chi)\bigr)\langle R_{66}\rangle + \gamma\bar n\langle R_{33}\rangle + \gamma(1+\bar n)  \nonumber \\
&\times&(1-\chi)\langle R_{88}\rangle - \frac{2\gamma\bar n}{3\sqrt{2}}(1-\chi)(\langle R_{32}\rangle  \nonumber \\
&+& \langle R_{23}\rangle), \nonumber \\
\langle \dot R_{77}\rangle &=&\frac{\gamma\bar n}{3}(1-\chi)(\langle R_{22}\rangle + 2\langle R_{33}\rangle) - 
\gamma\bigl(2-\chi  \nonumber \\
&+&\bar n(3-2\chi)\bigr)\langle R_{77}\rangle + \gamma\bar n\langle R_{44}\rangle + \gamma(1+\bar n)  \nonumber \\
&\times&(1-\chi)\langle R_{88}\rangle + \frac{2\gamma\bar n}{3\sqrt{2}}(1-\chi)(\langle R_{32}\rangle  \nonumber \\
&+& \langle R_{23}\rangle). 
\label{EqMa}
\end{eqnarray}
The equations of motion for the mean values of the off-diagonal elements, entering in Eqs.~(\ref{EqMa}), are given then as follows:
\begin{eqnarray}
\langle \dot R_{32}\rangle&=&\bigl\{3i\delta - \frac{\gamma}{2}\bigl(2 + \chi + 2\bar n(3 + \chi)\bigr)\bigr\}\langle R_{32}\rangle 
\nonumber \\
&-& \frac{\gamma}{2}(1+\bar n)\bigl\{\frac{4}{3\sqrt{2}}(1-\chi)\bigl(\langle R_{66}\rangle - \langle R_{77}\rangle\bigr) \nonumber \\
&-&\frac{2}{3}(1-\chi)\langle R_{56}\rangle + \frac{4}{3}(1+2\chi)\langle R_{65}\rangle \bigr\}, \nonumber \\
\langle \dot R_{65}\rangle&=&\bigl\{3i\delta - \frac{\gamma}{2}\bigl(4 + \chi + 2\bar n(3 + \chi)\bigr)\bigr\}\langle R_{65}\rangle 
 \nonumber \\
&-& \frac{\gamma}{2}\bar n\bigl\{\frac{4}{3\sqrt{2}}(1-\chi)\bigl(\langle R_{33}\rangle - \langle R_{44}\rangle\bigr) \nonumber \\
&-& \frac{2}{3}(1-\chi)\langle R_{23}\rangle + \frac{4}{3}(1+2\chi)\langle R_{32}\rangle \bigr\}, \label{offEq}
\end{eqnarray}
with $\langle R_{23}\rangle$=$[\langle R_{32}\rangle]^{\dagger}$ and $\langle R_{56}\rangle$=$[\langle R_{65}\rangle]^{\dagger}$. 

The steady-state solutions for the mean values of the involved atomic variables as well as the corresponding spatially scattered photon 
correlations functions are discussed in the next subsections, respectively.
\subsection{The steady-state solutions of the atomic variables}
Setting $\langle \dot R_{ij}\rangle \to 0$, $\{i,j \in 1, \cdots, 8\}$, in Eqs.~(\ref{EqMt}-\ref{offEq}) one arrives at the steady-state 
solutions for the involved atomic variables, as long as $\chi \not =1$, describing a small ensemble formed of three spatially 
equidistant placed two-level emitters interacting via their thermostat, i.e., 
\begin{eqnarray}
\langle R_{11}\rangle &=& \biggl(\frac{1+\bar n}{1+2\bar n} \biggr)^{3}, \nonumber \\
\langle R_{22}\rangle &=& \langle R_{33}\rangle=\langle R_{44}\rangle = \frac{\bar n(1+\bar n)^{2}}{(1+2\bar n)^{3}}, \nonumber \\
\langle R_{55}\rangle &=& \langle R_{66}\rangle=\langle R_{77}\rangle = \frac{\bar n^{2}(1+\bar n)}{(1+2\bar n)^{3}}, \nonumber \\
\langle R_{88}\rangle &=& \biggl(\frac{\bar n}{1+2\bar n} \biggr)^{3}, \label{sss}
\end{eqnarray}
while $\langle R_{32}\rangle = \langle R_{23}\rangle =0$ and $\langle R_{65}\rangle = \langle R_{56}\rangle =0$, respectively. 
Actually, the symmetrical as well as the anty-symmetrical three-atom collective states are populated as long as $\chi \not=1$. 
Moreover, the steady state solutions (\ref{sss}) do not depend on the collectivity, i.e. $\{\chi,\delta\}$, or on the single-atom 
spontaneous decay rate $\gamma$, regardless of the distance among the emitters. For weaker thermal baths, $\bar n \ll 1$,
one has that $\langle R_{11}\rangle \approx 1 - 3\bar n$ and $\langle R_{22}\rangle =\langle R_{33}\rangle=
\langle R_{44}\rangle \approx \bar n$, while the remained higher excited collective atomic states are inhibited, see 
Figure~(\ref{fig-2}). Bigger environmental temperatures, i.e. when $\bar n \gg 1$, lead to an identical population of each 
cooperative three-atom state which is equal to $1/8$, respectively.

However, if $\chi=1$ then from the master equation (\ref{MeqD}), or Eqs.~(\ref{EqMt}), follow the following steady-state 
solutions for the mean values of the populations in the symmetrical three-particle states:
\begin{eqnarray}
\langle R_{11}\rangle &=& \frac{(1+\bar n)^{3}}{(1+2\bar n)[(1+\bar n)^{2} + \bar n^{2}]}, \nonumber \\
\langle R_{22}\rangle &=& \frac{\bar n(1+\bar n)^{2}}{(1+2\bar n)[(1+\bar n)^{2} + \bar n^{2}]}, \nonumber \\
\langle R_{55}\rangle &=& \frac{\bar n^{2}(1+\bar n)}{(1+2\bar n)[(1+\bar n)^{2} + \bar n^{2}]}, \nonumber \\
\langle R_{88}\rangle &=& \frac{\bar n^{3}}{(1+2\bar n)[(1+\bar n)^{2} + \bar n^{2}]}.  \label{sDks}
\end{eqnarray}
The anti-symmetrical three-atom cooperative states are not populated since they decouple from the interaction with the 
thermal environment in this particular case, i.e. when $\chi=1$. Hence, $\langle R_{11}\rangle  + \langle R_{22}\rangle  +
 \langle R_{55}\rangle +\langle R_{88}\rangle=1$. Again, here, for lower surrounding temperatures, that is when $\bar n \ll 1$, 
one has that $\langle R_{11}\rangle \approx 1 - \bar n$ and $\langle R_{22}\rangle \approx \bar n$, whereas the steady-state 
populations of higher excited collective states $|5\rangle$ and $|8\rangle$ are too small. Stronger thermal heat baths, i.e. 
$\bar n \gg 1$, lead to $\langle R_{11}\rangle=\langle R_{22}\rangle =\langle R_{55}\rangle = \langle R_{88}\rangle = 1/4$ in 
the steady-state.

Generalising at this phase, regardless of the values of $\chi$, $0\le \chi \le 1$, at weaker thermal baths single-excitation 
collective states get populated preponderantly, whereas at stronger environmental heat reservoirs all the involved 
three-qubit cooperative states are equally populated, respectively. Actually, the cooperative three-qubit states are 
populated according to the Boltzmann distribution, see Exps.~(\ref{sss},\ref{sDks}).

In the following section, we shall use the steady-state populations, given by expressions (\ref{sss}) and (\ref{sDks}), in order 
to investigate the spatial photon correlations phenomena.
  
\subsection{Spatial photon correlation functions}
In the far-zone limit of experimental interest, the first- and the unnormalized second-order spatial photon correlation functions 
at position $\vec R$, with $R=|\vec R| \gg \lambda$, can be represented as follows, see e.g. \cite{gsag,wm,sczb,zficek,rewk,gb1,gb2},
\begin{eqnarray}
G_{1}(\vec R;t)=\sum^{N}_{j,l=1}\Phi_{R}(\vec r_{jl})\langle S^{+}_{j}(t)S^{-}_{l}(t)\rangle, \label{krf1} 
\end{eqnarray}
and
\begin{eqnarray}
G_{2}(\vec R_{1},\vec R_{2};t)&=&\sum^{N}_{j_{1},l_{1}=1}\sum^{N}_{j_{2},l_{2}=1}
\Phi_{R_{1}}(\vec r_{j_{1}l_{1}})\Phi_{R_{2}}(\vec r_{j_{2}l_{2}}) \nonumber\\
&\times&\langle S^{+}_{j_{1}}(t)S^{+}_{j_{2}}(t)S^{-}_{l_{2}}(t)S^{-}_{l_{1}}(t)\rangle, \label{krf2} 
\end{eqnarray}
whereas the unnormalized third-order spatial photon correlation function is given by the next expression \cite{zficek,gb1,gb2}, 
respectively,
\begin{eqnarray}
&{}&G_{3}(\vec R_{1},\vec R_{2}, \vec R_{3};t) \nonumber \\
&=&\sum^{N}_{j_{1},l_{1}=1}\sum^{N}_{j_{2},l_{2}=1}\sum^{N}_{j_{3},l_{3}=1}
\Phi_{R_{1}}(\vec r_{j_{1}l_{1}})\Phi_{R_{2}}(\vec r_{j_{2}l_{2}})\Phi_{R_{3}}(\vec r_{j_{3}l_{3}}) \nonumber\\
&\times&\langle S^{+}_{j_{1}}(t)S^{+}_{j_{2}}(t)S^{+}_{j_{3}}(t)S^{-}_{l_{3}}(t)S^{-}_{l_{2}}(t)S^{-}_{l_{1}}(t)\rangle. 
\label{krf3} 
\end{eqnarray}
Here, $\Phi_{R}(\vec r_{jl})=\Phi_{R}e^{i\frac{\omega_{0}}{c}\bigl(|\vec R -\vec r_{l}| - |\vec R - \vec r_{j}|\bigr)}$, while 
$\Phi_{R}=d^{2}\omega^{4}_{0}\bigl(1-\cos^{2}{\eta}\bigr)/\bigl(2c^{4}|\vec R -\vec r_{j}||\vec R - \vec r_{l}|\bigr)$ 
with $\eta$ being the angle between the direction of vector $\vec R$ and the dipole $\vec d$. 

Based on Exps.~(\ref{krf1}-\ref{krf3}), the normalized second-order photon-photon spatial correlation function is 
defined as \cite{gb1,gb2,zficek}:
\begin{eqnarray}
g^{(2)}(\vec R_{1},\vec R_{2};t) = \frac{G_{2}(\vec R_{1},\vec R_{2}; t)}{G_{1}(\vec R_{1};t)G_{1}(\vec R_{2};t)}, 
\label{nskrf}
\end{eqnarray}
whereas the normalised third-order spatial photon correlation function is given by the following expression 
\cite{zficek,gb1,gb2}:
\begin{eqnarray}
g^{(3)}(\vec R_{1},\vec R_{2},\vec R_{3};t) = \frac{G_{3}(\vec R_{1},\vec R_{2},\vec R_{3}; t)}
{G_{1}(\vec R_{1};t)G_{1}(\vec R_{2};t)G_{1}(\vec R_{3};t)}. \nonumber \\
\label{ntkrf}
\end{eqnarray}
Actually, we are interested in the steady state behaviours of the higher-order photon correlation functions, that is, of 
$g^{(2)}(0)=\lim_{t\to \infty}g^{(2)}(\vec R_{1},\vec R_{2};t)$ and 
$g^{(3)}(0)=\lim_{t\to \infty}g^{(3)}(\vec R_{1},\vec R_{2},\vec R_{3};t)$.

Now, using consecutively the relations (\ref{trSR}), which give the transformations from the single-atom operators to 
collective ones, and the steady-state solutions (\ref{sss}), one arrives at the following steady-state expressions for the 
spatial photon correlation functions when $N=3$, namely,
\begin{eqnarray}
G_{1}(\vec R)/\Phi_{R} = \frac{3\bar n}{1+2\bar n}, \label{skrf1} 
\end{eqnarray}
while
\begin{eqnarray}
G_{2}(\vec R_{1},\vec R_{2})/\kappa_{2} &=& 3 + \cos{\bigl[\frac{\omega_{0}r_{21}}{c}(\cos{\theta^{(1)}_{R_{2}}}
-\cos{\theta^{(1)}_{R_{1}}})\bigr]} \nonumber \\
&+&\cos{\bigl[\frac{\omega_{0}r_{31}}{c}(\cos{\theta^{(2)}_{R_{2}}} - \cos{\theta^{(2)}_{R_{1}}})\bigr]} \nonumber \\
&+& \cos{\bigl[\frac{\omega_{0}r_{32}}{c}(\cos{\theta^{(3)}_{R_{2}}} - \cos{\theta^{(3)}_{R_{1}}})\bigr]}, \label{skrf2} 
\end{eqnarray}
where $\kappa_{2}=2\Phi_{R_{1}}\Phi_{R_{2}}\bigl(\bar n/[1+2\bar n]\bigr)^{2}$ with $\Phi_{R_{\xi}} \propto R^{-2}_{\xi}$ 
and $\xi \in \{1,2\}$. Also, $\theta^{(\zeta)}_{R_{\xi}}$, $\zeta \in \{1,2,3\}$, are the angles between the inter particle 
vectors $\vec r_{jl}$, with $r_{jl}=|\vec r_{jl}| \equiv |\vec r_{j}-\vec r_{l}|$ and $\{j,l\} \in \{1,2,3\}$, and the detectors position 
vectors $\vec R_{\xi}$, $\xi \in \{1,2\}$, respectively. While obtaining the Exps.~(\ref{skrf1},\ref{skrf2}) we have also used 
that the steady-state mean values of all involved off-diagonal elements are zero as well as the corresponding second-order 
correlators, except for $\langle S^{+}_{1}S^{-}_{1}S^{+}_{2}S^{-}_{2}\rangle = \langle S^{+}_{1}S^{-}_{1}S^{+}_{3}S^{-}_{3}
\rangle =\langle S^{+}_{2}S^{-}_{2}S^{+}_{3}S^{-}_{3}\rangle \equiv (\bar n/[1+2\bar n])^{2}$. Notice that the first-order 
photon correlation function, i.e. $G_{1}(\vec R)$ or the photon scattered intensity, is an isotropic function of $\vec R$ and 
scales inversely proportional to the squared distance from the emitters to the detector. This is not the case anymore for 
higher-order photon correlation functions.

Correspondingly, the steady-state third-order spatial photon correlation function for the light scattered by thermally 
driven three-atom sample is,
\begin{eqnarray}
G_{3}(\vec R_{1},\vec R_{2},\vec R_{3})/\kappa_{3} &=& 3 + G_{3}(r_{21}) + G_{3}(r_{31}) + G_{3}(r_{32}) \nonumber \\
&+& \sum^{3}_{i=1}G^{(i)}_{3}. \label{skrf3} 
\end{eqnarray}
Here $\kappa_{3}=2\Phi_{R_{1}}\Phi_{R_{2}}\Phi_{R_{3}}\bigl(\bar n/[1+2\bar n]\bigr)^{3}$, while 
\begin{eqnarray*}
G_{3}(r_{21}) &=& \cos{\bigl[\frac{\omega_{0}r_{21}}{c}(\cos{\theta^{(1)}_{R_{2}}} - \cos{\theta^{(1)}_{R_{1}}})\bigr]}
\nonumber \\
&+&\cos{\bigl[\frac{\omega_{0}r_{21}}{c}(\cos{\theta^{(1)}_{R_{3}}} - \cos{\theta^{(1)}_{R_{1}}})\bigr]} \nonumber \\
&+& \cos{\bigl[\frac{\omega_{0}r_{21}}{c}(\cos{\theta^{(1)}_{R_{3}}}-\cos{\theta^{(1)}_{R_{2}}})\bigr]}, 
\end{eqnarray*}
\begin{eqnarray*}
G_{3}(r_{31}) &=& \cos{\bigl[\frac{\omega_{0}r_{31}}{c}(\cos{\theta^{(2)}_{R_{2}}} - \cos{\theta^{(2)}_{R_{1}}})\bigr]}
\nonumber \\
&+&\cos{\bigl[\frac{\omega_{0}r_{31}}{c}(\cos{\theta^{(2)}_{R_{3}}} - \cos{\theta^{(2)}_{R_{1}}})\bigr]} \nonumber \\
&+& \cos{\bigl[\frac{\omega_{0}r_{31}}{c}(\cos{\theta^{(2)}_{R_{3}}} - \cos{\theta^{(2)}_{R_{2}}})\bigr]}, 
\end{eqnarray*}
and
\begin{eqnarray*}
G_{3}(r_{32}) &=& \cos{\bigl[\frac{\omega_{0}r_{32}}{c}(\cos{\theta^{(3)}_{R_{2}}} - \cos{\theta^{(3)}_{R_{1}}})\bigr]}
\nonumber \\
&+&\cos{\bigl[\frac{\omega_{0}r_{32}}{c}(\cos{\theta^{(3)}_{R_{3}}} - \cos{\theta^{(3)}_{R_{1}}})\bigr]} \nonumber \\
&+& \cos{\bigl[\frac{\omega_{0}r_{32}}{c}(\cos{\theta^{(3)}_{R_{3}}} - \cos{\theta^{(3)}_{R_{2}}})\bigr]}, 
\end{eqnarray*}
respectively, whereas,
\begin{eqnarray*}
G^{(1)}_{3}&=& \cos{\bigl[\frac{\omega_{0}}{c}(r_{21}\cos{\theta^{(1)}_{R_{1}}} + r_{32}\cos{\theta^{(3)}_{R_{3}}} 
- r_{31}\cos{\theta^{(2)}_{R_{2}}})\bigr]} \nonumber \\
&+& \cos{\bigl[\frac{\omega_{0}}{c}(r_{21}\cos{\theta^{(1)}_{R_{1}}} + r_{32}\cos{\theta^{(3)}_{R_{2}}} 
- r_{31}\cos{\theta^{(2)}_{R_{3}}})\bigr]},
\end{eqnarray*}
\begin{eqnarray*}
G^{(2)}_{3}&=& \cos{\bigl[\frac{\omega_{0}}{c}(r_{21}\cos{\theta^{(1)}_{R_{2}}} + r_{32}\cos{\theta^{(3)}_{R_{3}}} 
- r_{31}\cos{\theta^{(2)}_{R_{1}}})\bigr]} \nonumber \\
&+& \cos{\bigl[\frac{\omega_{0}}{c}(r_{21}\cos{\theta^{(1)}_{R_{2}}} + r_{32}\cos{\theta^{(3)}_{R_{1}}} 
- r_{31}\cos{\theta^{(2)}_{R_{3}}})\bigr]},
\end{eqnarray*}
and
\begin{eqnarray*}
G^{(3)}_{3}&=& \cos{\bigl[\frac{\omega_{0}}{c}(r_{21}\cos{\theta^{(1)}_{R_{3}}} + r_{32}\cos{\theta^{(3)}_{R_{2}}} 
- r_{31}\cos{\theta^{(2)}_{R_{1}}})\bigr]} \nonumber \\
&+& \cos{\bigl[\frac{\omega_{0}}{c}(r_{21}\cos{\theta^{(1)}_{R_{3}}} + r_{32}\cos{\theta^{(3)}_{R_{1}}} 
- r_{31}\cos{\theta^{(2)}_{R_{2}}})\bigr]},
\end{eqnarray*}
where we have used that the only non-zero three-particle correlator is 
$\langle S^{+}_{1}S^{+}_{2}S^{+}_{3}S^{-}_{3}S^{-}_{2}S^{-}_{1}\rangle \equiv \langle R_{88}\rangle$.
Also, here, one has that $r_{21}=r_{31}=r_{32} \equiv r$, with $R \gg r$. 

In the following section, we shall describe the spatial interference effects of the photons scattered by the considered 
three-atom sample, mutually interacting via their common thermostat.  Notice that, $g^{(2)}(0)<1$ characterizes 
sub-Poissonian, $g^{(2)}(0)>1$ super-Poissonian, and $g^{(2)}(0)=1$ Poissonian photon statistics. Furthermore,
the condition $g^{(3)}(0) < g^{(2)}(0) <1$ means that the scattered photon beam consists of non-classical single 
photons.

\section{Results and Discussion \label{RD}}
We proceed in the following to describe the spatial interference features based on the second- and third-order photon 
correlation functions, respectively. This is achieved via photon detection of the spontaneously scattered electromagnetic 
field by the considered three-atom sample, arranged in an equilateral triangle geometrical configuration, see Figure~(\ref{fig-1}). 
In this respect, from Exps.~(\ref{skrf1},\ref{skrf2}) and Exp.~(\ref{nskrf}) immediately follows the normalised second-order spatial 
photon correlation function, namely,
\begin{eqnarray}
g^{(2)}(0)&=&\frac{2}{9}\biggl\{3 + \cos{\bigl[\frac{\omega_{0}r_{21}}{c}(\cos{\theta^{(1)}_{R_{2}}}
-\cos{\theta^{(1)}_{R_{1}}})\bigr]} \nonumber \\
&+&\cos{\bigl[\frac{\omega_{0}r_{31}}{c}(\cos{\theta^{(2)}_{R_{2}}} - \cos{\theta^{(2)}_{R_{1}}})\bigr]} \nonumber \\
&+& \cos{\bigl[\frac{\omega_{0}r_{32}}{c}(\cos{\theta^{(3)}_{R_{2}}} - \cos{\theta^{(3)}_{R_{1}}})\bigr]}\biggr\}. \label{g2krf} 
\end{eqnarray}
One can easily observe that for a two-photon detector one always has that $g^{(2)}(0) = 4/3$. The same result persists 
if the two detectors are oppositely placed along a line, normal to the atomic plane, which passes through the triangle's 
centre. If, for instance, the two detectors, $\{D_{1},D_{2}\}$, are arranges as depicted in Figure (\ref{fig-1}), while the third 
one, i.e. $D_{3}$, is being omitted then the spatial second-order photon-photon correlation function becomes
\begin{eqnarray}
g^{(2)}(0)=\frac{2}{9}\bigl\{3 + \cos{[\sqrt{3}\omega_{0}r/c]} + 2\cos{[\sqrt{3}\omega_{0}r/2c]}\bigr\}, \nonumber \\
\label{g2kp} 
\end{eqnarray}
which follows from Exp.~(\ref{g2krf}). If $\sqrt{3}\omega_{0}r/(2c) = (\pi/2)m$, $m \in \{1, 2, 3, \cdots \}$, then $g^{(2)}(0)$, 
given by Exp.~(\ref{g2kp}), varies between $g^{(2)}(0)=4/9$ and $g^{(2)}(0)=4/3$, i.e., one has sub- to super-Poissonian 
photon statistics, respectively, see also the solid line in Figure~(\ref{fig-3}). Similar result is obtained when the two detectors 
are situated in the atomic triangle's plane and oppositely fixed along the vector $\vec r_{21}$, resulting in the following 
expression for the spatial second-order photon correlation function, namely,
\begin{eqnarray}
g^{(2)}(0)=\frac{2}{9}\bigl\{3 + \cos{[2\omega_{0}r/c]} + 2\cos{[\omega_{0}r/c]}\bigr\}. \label{g2kfp} 
\end{eqnarray}
Now, if $\omega_{0}r/c = (\pi/2)m$, $m \in \{1, 2, 3, \cdots \}$, then $g^{(2)}(0)$ lies between $g^{(2)}(0)=4/9$ and 
$g^{(2)}(0)=4/3$, too. The first minimum of $g^{(2)}(0)$ in Figure~(\ref{fig-3}), equal to $g^{(2)}_{min}(0)=1/3$, occurs 
at $x/\pi=2/3$. Thus, generally, depending on the distances among the emitters and detector positions, the scattered 
photon statistics possesses quantum or close to coherent photon statistics features. Furthermore, if the two detectors 
are placed in symmetric positions, sub-wavelength interference fringes can be observed, see e.g. Exp.~(\ref{g2kfp}) 
and the solid line in Figure~(\ref{fig-3}), where one should replace $x$ with $x \to \omega_{0}r/c$.

In the same vain, from Exp.~(\ref{skrf1}) and Exp.~(\ref{skrf3}), one obtains the expression for the normalized third-order 
spatial photon correlation function, i.e.,
\begin{eqnarray}
g^{(3)}(0) &=&\frac{2}{27}\bigl\{ 3 + G_{3}(r_{21}) + G_{3}(r_{31}) + G_{3}(r_{32}) \nonumber \\
&+& \sum^{3}_{i=1}G^{(i)}_{3}\bigr\}. \label{g3krf} 
\end{eqnarray}
Particularly, for a single three-photon detector, from Exp.~(\ref{g3krf}) follows that:
\begin{eqnarray}
g^{(3)}(0) =\frac{4}{9}\bigl\{2 + \cos{\bigl[\frac{\omega_{0}}{cR}(\vec r_{21}+\vec r_{32} - \vec r_{31}, \vec R)\bigr]}\bigr\}. 
\label{g3tr} 
\end{eqnarray}
Since in our case $\vec r_{21}+\vec r_{32} = \vec r_{31}$, see Figure~(\ref{fig-1}), then one obtains that $g^{(3)}(0)=4/3$. 
Thus, generalizing at this stage, the probabilities to detect simultaneously a two- or a three-photon event, using a single 
two- or three-photon detector, are equal, i.e. $g^{(2)}(0)=g^{(3)}(0)=4/3$, while this is being an isotropic process. In the 
case when three distinct photon detectors, $\{D_{1},D_{2},D_{3}\}$, are arranged as shown in Figure~(\ref{fig-1}), then the 
spatial third-order correlation function takes the form:
\begin{figure}[t]
\includegraphics[width =7cm]{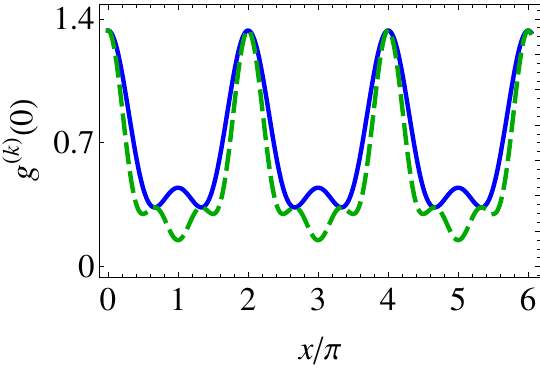}
\caption{\label{fig-3}
The steady-state higher-order photon correlation functions $g^{(k)}(0)$ versus $x/\pi$, where $x=\sqrt{3}\omega_{0}r/(2c)$, 
based on Exp.~(\ref{g2kp}) and Exp.~(\ref{g3kp}), respectively. The solid line depicts the second-order photon  
correlation function, i.e. $k=2$, whereas the long-dashed curve stands for the third-order photon correlation function, where $k=3$.}
\end{figure}
\begin{eqnarray}
g^{(3)}(0) &=& \frac{2}{27}\bigl\{7 + 3\cos{[\sqrt{3}\omega_{0}r/c]} + 6\cos{[\sqrt{3}\omega_{0}r/2c]} \nonumber \\
&+& 2\cos{[3\sqrt{3}\omega_{0}r/2c]}\bigr\}. \label{g3kp} 
\end{eqnarray}
If, for instance, $\sqrt{3}\omega_{0}r/(2c) = (\pi/2)m$, $m \in \{1, 2, 3, \cdots \}$, then $g^{(3)}(0)$ ranges between 
$g^{(3)}(0)=4/27$ to $g^{(3)}(0)=4/3$, see the long-dashed curve in Figure~(3). Comparing with the results for the 
second-order correlation function, given by Exp.~(\ref{g2kp}), one observes that when $\sqrt{3}\omega_{0}r/(2c) = \pi/2$, 
the value for $g^{(3)}(0)=8/27$ is lower than that of $g^{(2)}(0)=4/9$, while $g^{(2)}(0) < 1$, see also Figure~(3). This 
indicates that streams of single photons, possessing sub-Poissonian statistics, is generating when both detectors, i.e. 
$\{D_{1},D_{2}\}$, click, see Figure~(\ref{fig-1}). Similar result occurs when $\sqrt{3}\omega_{0}r/(2c) = 19\pi/2$, for 
instance. The difference here is that if $\sqrt{3}\omega_{0}r/(2c) = \pi/2$, then $r/\lambda \approx 0.3$, whereas when 
$\sqrt{3}\omega_{0}r/(2c) = 19\pi/2$, then $r/\lambda \approx 5.5$, respectively. This means that in the first case we are 
in the collective regime of interaction, while in the second case not. One may expect then that the single-photon emission 
processes, based on the condition $g^{(3)}(0) < g^{(2)}(0)< 1$, might taking place because of the dipole-dipole blockade 
effect occurring at smaller inter-particle distances, $r/\lambda < 1$. At larger atomic intervals, that is when $r/\lambda >1$, 
the single-photon emission processes are due to spatial interference phenomena, respectively. Furthermore, if 
$\chi \not = 1$, the normalised spatial photon correlation functions do not explicitly depend on the thermostat's 
properties,  although the unnormalised photon correlation functions, i.e. $\{G_{1},G_{2},G_{3}\}$, indeed depend 
explicitly on the mean thermal photon number $\bar n$, see Exps.~(\ref{skrf1}-\ref{skrf3}).

In order to elucidate if this is indeed the case, namely, the sub-Poissonian photon statistics may occurring due to the 
dipole-dipole blockade or high-order spatial interference, or something else, we shall focus next on the Dicke model 
for which $\chi=1$. In this respect, the unnormalised photon correlation functions are given as: 
$G_{1} \sim \langle S^{+}S^{-}\rangle$, $G_{2}\sim \langle S^{+}S^{+}S^{-}S^{-}\rangle$ and 
$G_{3}\sim \langle S^{+}S^{+}S^{+}S^{-}S^{-}S^{-}\rangle$, respectively. Correspondingly, the normalised second- 
and third-order photon correlation functions are defined as follows:
\begin{eqnarray}
g^{(2)}(0) = \frac{\langle S^{+}S^{+}S^{-}S^{-}\rangle}{\langle S^{+}S^{-}\rangle^{2}}, 
\label{g2kfD} 
\end{eqnarray}
and
\begin{eqnarray}
g^{(3)}(0) = \frac{\langle S^{+}S^{+}S^{+}S^{-}S^{-}S^{-}\rangle}{\langle S^{+}S^{-}\rangle^{3}}, 
\label{g3kfD} 
\end{eqnarray}
where $S^{+}=S^{+}_{1} + S^{+}_{2} +S^{+}_{3}$ while $S^{-}=[S^{+}]^{\dagger}$. Inserting,
\begin{eqnarray*}
S^{+} &=& \sqrt{3}(R_{21}+R_{85}) + 2R_{52}, ~~ {\rm and}~ \nonumber \\
S^{-} &=& \sqrt{3}(R_{12}+R_{58}) + 2R_{25},
\end{eqnarray*}
which follow from (\ref{trSR}) in the Dicke limit, into relations (\ref{g2kfD}) and (\ref{g3kfD}) and using the steady-state 
solutions (\ref{sDks}), one arrives at the following normalised expressions for the photon correlation functions in the 
Dicke limit, namely,
\begin{eqnarray}
g^{(2)}(0) = \frac{12(1+2\bar n)^{2}[(1+\bar n)^{2} +\bar n^{2}]}{[10\bar n^{2} + 10\bar n +3]^{2}}, \label{g2D} 
\end{eqnarray}
and
\begin{eqnarray}
g^{(3)}(0) = \frac{36(1+2\bar n)^{2}[(1+\bar n)^{2} +\bar n^{2}]^{2}}{[10\bar n^{2} + 10\bar n +3]^{3}}. \label{g3D} 
\end{eqnarray}
Their ratio is
\begin{eqnarray}
g^{(2)}(0)/g^{(3)}(0) = \frac{3 + 10\bar n + 10\bar n^{2}}{3[(1+\bar n)^{2} + \bar n^{2}]}. \label{rapg}
\end{eqnarray}
Respectively, the scattered intensity by the three-atom sample in the Dicke limit is
\begin{eqnarray}
G_{1}/\Phi_{R} = \frac{\bar n(3 + 10\bar n + 10\bar n^{2})}{(1+2\bar n)\bigl((1+\bar n)^{2} + \bar n^{2}\bigr)}. 
\label{G1Dk}
\end{eqnarray}
Contrary to the spatially extended three-atom sample where $\chi \not=1$, in the Dicke limit the normalised second- 
and third-order photon correlation functions do depend on the thermostat's properties, that is on $\bar n$, see 
Exps.~(\ref{g2D},\ref{g3D}). Particularly, if $\hbar\omega_{0}/(k_{B}T) \gg 1$, then $g^{(2)}(0)=g^{(3)}(0) =4/3$. 
However, when $\hbar\omega_{0}/(k_{B}T) \ll 1$, one has that $g^{(2)}(0)=24/25  <1$ and $g^{(3)}(0)=144/250$, 
while their ratio is $g^{(2)}(0)/g^{(3)}(0) = 10/6 >1$. One can observe that in the Dicke limit with higher environmental 
temperatures, i.e. $\bar n \gg 1$, one has that $g^{(3)}(0) < g^{(2)}(0) <1$ which is similar to the photon blockade 
effect occurring due to dipole-dipole interactions. However, the photon correlation functions do not depend on the 
dipole-dipole coupling strength, $\delta$, and so does the populations of the involved three-atom collective states. 
Furthermore, stronger surrounding heat reservoirs lead here to equal population of all involved three-qubit cooperative 
states and not preponderantly of the single-excitation cooperative state, i.e. $|2\rangle$. Hence, on the cooperative 
atomic transitions $|8\rangle \to |5\rangle \to |2\rangle \to |1\rangle$ streams of single photons are emitted with 
sub-Poissonian statistics. Larger atomic ensembles, however, will not lead to photon emissions with quantum 
features \cite{hss}, unless $\delta$ approaches $\omega_{0}$, see Ref.~\cite{mma}. Therefore, the generation of 
photon fluxes by considered three-atom sample, consisting of single photons which possess quantum photon 
statistics, occurs due to few-emitters-thermostat interaction's nature and not because of the dipole-dipole 
blockade phenomena.

Thus, generalizing, depending on the mutual spatial separations among the emitters, a photon flux with sub-Poissonian 
statistics is generated. However, the physics behind is different, namely, at shorter inter particle distances the 
few-atoms-thermostat interaction's nature is responsible for photon emissions with quantum properties, 
whereas at larger distances - the spatial interference phenomena. Furthermore, if the two detectors are placed 
in symmetric positions, sub-wavelength interference fringes can be observed too. In the Dicke limit, the photon 
emission processes are isotropic, which is not the case anymore for higher-order photon correlations in an 
extended three-atom sample.  

Finally, there is not a smooth transition between an extended atomic sample, where $\chi \not =1$, and the Dicke 
model with $\chi=1$. The reason is that even when $\chi \to 1$, the anti-symmetrical cooperative three-atom states 
are populated in the steady-state according to Exps.~(\ref{sss}). The population time of the anti-symmetrical states 
then is slow, proportional to $t \sim [\gamma(1-\chi)]^{-1}$. So, at time-intervals shorter than the population time 
of the anti-symmetrical states, the atomic sample will behave based on the Dicke model predictions if the inter 
particle space intervals are short enough compared to the photon emission wavelength, $\lambda$. Note also 
that if $T=0$, i.e. $\bar n=0,$ then $\{G_{1}, G_{2}, G_{3}\}=0$ too, and the normalized second- or third-order 
photon correlation functions are inappropriate.

\section{Summary \label{sum}}
We have investigated the spatial photon interference phenomena based on the higher order photon correlation 
functions, while focusing on extended or Dicke-like three-atom sample. The two-level emitters are arranged 
at the vertices of an equilateral triangle and dipole-dipole interact through their environmental thermostat. 
As a consequence, we demonstrated that the employed atomic sample spontaneously generates streams of 
single-photons, possessing sub-Poissonian photon statistics. Furthermore, at smaller inter particle distances 
the effect is due to the thermostat-few-emitters interaction's nature, whereas at larger inter atomic distances 
because of high-order spatial photon interference phenomena, respectively. If the two detectors are placed 
in symmetric positions, sub-wavelength interference fringes can be observed as well. The photon emission 
process is isotropic within the Dicke limit and spatially dependent in the case of an extended atomic model.

\acknowledgments
The financial support from The Ministry of Education and Research, via grants No. 25.80012.5007.73SE and No. 011205, are 
gratefully acknowledged.


\end{document}

When the two detectors lye in the atomic triangle's plane and oppositely fixed along the vector $\vec r_{21}$, while the third one is placed 
along a line normal to the triangle's centre, then the third-order spatial photon correlation function is given by the next expression:
\begin{eqnarray}
g^{(3)}(0) &=& \frac{2}{27}\bigl\{5 + \cos{[2\omega_{0}r/c]} + 4(\cos{[\omega_{0}r/2c]} \nonumber \\
&+& \cos{[\omega_{0}r/c]} + \cos{[3\omega_{0}r/2c]})\bigr\}. \label{g3kfp} 
\end{eqnarray}
If $\omega_{0}r/c=\pi m$, $m \in \{1, 2, \cdots \}$, then $g^{(3)}(0)$, given by Exp.~(\ref{g3kfp}), may vary between 
$g^{(3)}(0)=4/27$ and $g^{(3)}(0)=4/3$.